%% file: main.tex
\documentclass[review,3p]{elsarticle}

\usepackage{graphicx}
\usepackage{epsfig}
\usepackage{color}
\usepackage{psfrag}
\usepackage{subfigure}
\usepackage{subfigmat}
\usepackage{subeqnarray}
\usepackage{amsmath}
\usepackage{amsfonts,stmaryrd,amssymb,theorem}
\usepackage{xcolor,colortbl}
\usepackage[mathscr]{euscript}
\usepackage[ruled]{algorithm2e}
\usepackage{algorithmic}
\usepackage{changepage}
\usepackage{multirow}
\graphicspath{{../Figures/}{./Figure/}}
\usepackage{rotating}
\usepackage{bm}
\usepackage{lineno}
\usepackage{textgreek}
\usepackage{enumitem}
\usepackage{array}
\usepackage{tikz}
\usepackage{booktabs}

\usepackage[export]{adjustbox}
\usepackage{epstopdf}
\usepackage[capitalize]{cleveref}

\input{abbrevs}

\journal{AIAA SciTech 2018, Kissimmee, FL}

\begin{document}

\title{Direct numerical simulations of turbulent channel flow under transcritical conditions}


\author[Stanford]{Peter C. Ma}
\author[ctr]{Xiang I. A. Yang\corref{cor}}
\ead{xiangyang@stanford.edu}
\author[Stanford]{Matthias Ihme}

\address[Stanford]{Department of Mechanical Engineering, Stanford University, Stanford, CA 94305, USA}
\address[ctr]{Center for Turbulence Research, Stanford University, Stanford, CA 94305, USA}

\cortext[cor]{Corresponding author}

\begin{abstract}
Turbulent flows under transcritical conditions are present in regenerative cooling systems of rocker engines and extraction processes in chemical engineering. 
The turbulent flows and the corresponding heat transfer phenomena in these complex processes are still not well understood experimentally and numerically. 
The objective of this work is to investigate the turbulent flows under transcritical conditions using DNS of turbulent channel flows. 
A fully compressible solver is used in conjunction with a Peng-Robinson real-fluid equation of state to describe the transcritical flows. 
A channel flow with two isothermal walls is simulated with one heated and one cooled boundary layers. 
The grid resolution adopted in this study is slightly finer than that required for DNS of incompressible channel flows.
The simulations are conducted using both fully (FC) and quasi-conservative (QC) schemes to assess their performance for transcritical wall-bounded flows.
The instantaneous flows and the statistics are analyzed and compared with the canonical theories. 
It is found that results from both FC and QC schemes qualitatively agree well with noticeable difference near the top heated wall, where spurious oscillations in velocity can be observed.
Using the DNS data, we then examine the usefulness of Townsend attached eddy hypothesis in the context of flows at transcritical conditions. 
It is shown that the streamwise energy spectrum exhibits the inverse wavenumber scaling and that the streamwise velocity structure function follows a logarithmic scaling, thus providing support to the attached eddy model at transcritical conditions.

\end{abstract}

\maketitle

\section{Introduction}
Transcritical turbulent flows are present in regenerative cooling systems of rocker engines where a cryogenic fuel in a supercritical state flows through wall-embedded channels to absorb the heat from combustion chamber. 
Supercritical fluids are also used as coolants in nuclear reactors~\cite{yoo2013turbulent}. 
In chemical engineering, supercritical fluids are commonly used for extraction processes and a recent review paper is given by Brunner~\cite{brunner2010applications}. 
At elevated pressures, the mixture properties exhibit liquid-like densities and gas-like diffusivities, and the surface tension and enthalpy of vaporization approach zero~\cite{yang2000modeling}. 
This phenomenon was shown by recent experimental studies~\cite{oschwald2006injection, mayer2000injection, chehroudi2012recent}. 
Large thermo-transport gradients are present under transcritical conditions when the fluid goes through the pseudo-boiling process~\cite{chehroudi2012recent, banuti2015crossing, Banuti2016sub, banuti2017seven, raju2017widom}. 
Specifically, the density ratio between the dense liquid and the light gaseous fluid can be on the order of $\mathcal{O}$(100). 
A local peak of the specific heat is present across the pseudo-boiling point which resembles the subcritical phase change phenomenon. 
The large thermodynamic variations along with transport anomalies alter the classical behaviors of turbulent flows and the corresponding heat transfer processes. 
However, turbulent flows in these complex processes under transcritical conditions are still not well understood experimentally and numerically.

Several previous studies use direct numerical simulations (DNS) for the study of transcritical turbulent flows. 
Fully compressible DNS of turbulent shear layers with supercritical fluids are investigated by Miller~et~al.~\cite{miller1999direct, miller2001direct} and Okong'o~and~Bellan~\cite{okong2002direct}. Bae~et~al.~\cite{bae2005direct, bae2008direct} used a low-Mach solver to study the turbulent heat transfer of CO$_2$ at supercritical pressure flowing in heated vertical tubes at several different Reynolds numbers and operating conditions. 
The heat transfer of supercritical flows under similar configurations are investigated by Nemati~et~al.~\cite{nemati2015mean, peeters2016turbulence} and Peeters~et~al.~\cite{peeters2016turbulence}. 
The effect of temperature dependent density and viscosity on turbulence in channel flows are investigated theoretically and numerically to assess the validity of the semi-local scaling by Patel~et~al.~\cite{patel2015semi}. 
Kawai~et~al.~\cite{kawai2015robust, kawai2016direct} used a fully compressible solver to conduct a DNS of zero-pressure-gradient heated transcritical turbulent boundary layers on a flat plate at supercritical pressure conditions. 
Kim~et~al.~\cite{kim2017numerical} carried out turbulent channel flow simulations of supercritical R-134a using Peng-Robinson equation of state to describe the thermodynamics.

To study transcritical flows numerically, diffuse-interface methods have been widely used and different numerical schemes have been adopted~\cite{schmitt2010large, ruiz2012unsteady, terashima2012approach, matheis2016multi, hickey2013large, kawai2015robust, lacaze2017robust, pantano2017oscillation, yang2017comparison, unnikrishnan2017subgrid}. Traditionally, fully conservative (FC) schemes have been used for transcritical flows. However, several groups reported numerical difficulties or even failures with FC schemes in conjunction with a real-fluid state equation, due to the occurrence of spurious pressure oscillations~\cite{schmitt2010large, matheis2016multi, hickey2013large, lacaze2017robust}. This has motivated the development of quasi-conservative (QC) schemes for transcritical flows. Schmitt~et~al.~\cite{schmitt2010large, ruiz2012unsteady} added a correction term in the energy equation by connecting the artificial dissipation terms in the mass, momentum, and energy conservation equations and setting the pressure differential to zero. Since the correction term is not in flux form, the scheme is not strictly energy-conservative. Terashima, Kawai and coworkers~\cite{terashima2012approach, kawai2015robust} solved a transport equation for pressure instead of the total energy equation in their finite difference solver so that the pressure equilibrium across contact interfaces is maintained. Pantano~et~al.~\cite{pantano2017oscillation} formulated a numerical scheme for transcritical contact and shock problems, which introduces an additional non-conservative transport equation for maintaining the mechanical equilibrium of pressure. Ma~et~al.~\cite{ma2014supercritical, ma2017framework, MaJCP2017} extended a double-flux model to the transcritical regime, which is capable to eliminate spurious pressure oscillations. The comparison between FC and QC schemes for studying turbulent wall-bounded flows is currently limited in literature.

To date, low-cost computational fluid dynamic (CFD) models (e.g. Reynolds-averaged-Navier-Stokes) have not been able to accurately predict the wall heat transfer rates at transcritical and supercritical conditions because of an insufficient description of the near-wall turbulent heat transfer \cite{yoo2013turbulent}.
While efforts have been made to enable wall-modeled large-eddy simulation (WMLES) capabilities for problems involving wall heat transfer~\cite{kawai2013dynamic, yang2016heat, ma2016development, ma2017flamelet, ma2017non}, applications of this cost-efficient tool has so far been limited to ideal-gas flows.
The objective of this work is to investigate the turbulent flows under transcritical conditions using DNS of turbulent channel flows, and attempt to use DNS to inform low-cost CFD models by examining commonly employed physical models for variable property fluids. 

The remainder of the paper is organized as follows.
Mathematical formulation including governing equations, thermodynamic and transport property evaluation, numerical methods used in this study, and the details of the computational configurations is presented in \cref{sec:math}. In \cref{sec:results}, the DNS results are presented and the performance of different schemes is assessed. The attached eddy model~\cite{townsend1980structure, woodcock2015statistical, yang2016hierarchical} is then examined for transcritical flows. The paper finishes with conclusions in \cref{sec:conclusions}.

\section{Mathematical Formulation}\label{sec:math}

\subsection{Governing equations}
The governing equations for the description of the transcritical flows studied here are the conservation of mass, momentum, and total energy, which take the following form
\begin{subequations}
    \label{eqn:governingEqn}
    \begin{align}
        \frac{\partial \rho}{\partial t} +\nabla \cdot (\rho \boldsymbol{u}) &= 0\,,\\
        \frac{\partial (\rho \boldsymbol{u})}{\partial t} + \nabla \cdot (\rho \boldsymbol{u} \boldsymbol{u} + p \boldsymbol{I}) &= \nabla \cdot \boldsymbol{\tau} + \boldsymbol{f} \,,\\
        \frac{\partial  (\rho E)}{\partial t} + \nabla \cdot [\boldsymbol{u}(\rho E + p)] &= \nabla \cdot (\boldsymbol{\tau} \cdot \boldsymbol{u}) -\nabla \cdot \boldsymbol{q} + \boldsymbol{u} \cdot \boldsymbol{f} \,,
    \end{align}
\end{subequations}
where $\rho$ is the density, $\boldsymbol{u}$ is the velocity vector, $p$ is the pressure, $\boldsymbol{f}$ is the body force, and $E$ is the specific total energy. The viscous stress tensor and heat flux are written as
\begin{subequations}
    \begin{align}
        &\boldsymbol{\tau} =  \mu \left[ \nabla \boldsymbol{u} + (\nabla \boldsymbol{u})^T \right] -\frac{2}{3}\mu (\nabla \cdot \boldsymbol{u}) \boldsymbol{I} \;,\\
        &\boldsymbol{q} = - \lambda \nabla T\;,
    \end{align}
\end{subequations}
where $T$ is the temperature, $\mu$ is the dynamic viscosity, and $\lambda$ is the thermal conductivity. The total energy is related to the internal energy and the kinetic energy
\begin{equation}
    \rho E = \rho e + \frac{1}{2}\rho \boldsymbol{u} \cdot \boldsymbol{u}\,.
\end{equation}
The system is closed with an equation of state (EoS), which is here written in pressure-explicit form as
\begin{equation}
    \label{eqn:eos}
    p = f(\rho, e)\,,
\end{equation}
and this will be discussed in detail in the next subsection.

\subsection{Thermodynamic relations}\label{sec:eos}
The Peng-Robinson (PR) cubic equation of state (EoS)~\cite{poling2001properties, peng1976new} is used in this study due to its reasonable accuracy, computational efficiency, and prevailing usage, which can be written as
\begin{equation}
    \label{eqn:preos}
    p = \frac{R T}{v - b} - \frac{a\alpha}{v^2+2bv-b^2}\,,
\end{equation}
where $R$ is the gas constant, $v = 1/\rho$ is the specific volume, and the parameters $a\alpha$ and $b$ are dependent on temperature and composition to account for effects of intermolecular forces. The parameters $a$ and $b$ are evaluated as
\begin{subequations}
    \label{EQ_PR_EQUATION_alphabeta}
    \begin{align}
        a &= 0.457236\frac{R^2 T_{c}^2}{p_{c}}\;,\\
        b &= 0.077796 \frac{R T_{c}}{p_{c}}\;,\\
        \alpha &= \left[1+c\left(1-\sqrt{\frac{T}{T_{c}}}\right)\right]^2\;,
    \end{align}
\end{subequations}
where $T_{c}$ and $p_{c}$ are the critical temperature and pressure and
\begin{equation}
    c = 0.37464 + 1.54226\omega - 0.26992\omega^2\,,
\end{equation}
where $\omega$ is the acentric factor.

For general real fluids, thermodynamic quantities are typically evaluated from the ideal-gas value plus a departure function that accounts for the deviation from the ideal-gas behavior. The ideal-gas enthalpy, entropy and specific heat can be evaluated from the commonly used NASA polynomials which have a reference temperature of 298~K.
The specific internal energy can be written as,
\begin{equation}
    \label{eqn:generalInternalEnergy}
    e(T, \rho) = e^{\text{ig}}(T) + \int_v^{+\infty} \left[p - T\left(\frac{\partial p}{\partial T}\right)_{v}\right] dv\,,
\end{equation}
where superscript ``ig" indicates the ideal-gas value of the thermodynamic quantity, and \cref{eqn:generalInternalEnergy} can be integrated analytically for PR EoS:
\begin{equation}
    \label{eqn:internalEnergy}
    e = e^{\text{ig}} + K_1\left(a\alpha - T\frac{\partial a\alpha}{\partial T}\right),
\end{equation}
in which $K_1$ is defined as
\begin{equation}
    K_1 = \int_{+\infty}^v \frac{1}{v^2+2bv-b^2}dv = \frac{1}{\sqrt{8} b}\ln\left(\frac{v+(1-\sqrt{2})b}{v+(1+\sqrt{2})b} \right) \label{eqn:K1}\,,
\end{equation}
and
\begin{equation}
    \frac{\partial a\alpha}{\partial T} = - a c \sqrt{\frac{\alpha}{TT_c}}\;.
\end{equation}
The specific enthalpy can be evaluated from the thermodynamic relation $h = e + pv$, and we have
\begin{equation}
    h = h^\text{ig} - R T + K_1\left[a\alpha - T \frac{\partial a\alpha}{\partial T}\right] + pv\,.
\end{equation}

The specific heat capacity at constant volume is evaluated as
\begin{equation}
    c_v = \left(\frac{\partial e}{\partial T}\right)_{v} = c^\text{ig}_v - K_1 T \frac{\partial^2 a\alpha}{\partial T^2}\,,
\end{equation}
and the specific heat capacity at constant pressure is evaluated as
\begin{equation}
    c_p = \left(\frac{\partial h}{\partial T}\right)_{p} = c^\text{ig}_p - R - K_1 T \frac{\partial^2 a\alpha}{\partial T^2} - T \frac{(\partial p/\partial T)^2_{v}}{(\partial p/\partial v)_{T}}\,,
\end{equation}
where
\begin{equation}
    \frac{\partial^2 a\alpha}{\partial T^2} = \frac{ac^2}{2TT_c} + \frac{ac}{2} \sqrt{\frac{\alpha}{T^3T_c}}\,,
\end{equation}
and
\begin{subequations}
    \begin{align}
        \left(\frac{\partial p}{\partial T}\right)_{v} &= \frac{R}{v-b} - \frac{\partial a\alpha/\partial T}{v^2+2bv-b^2}\;, \\
        \left(\frac{\partial p}{\partial v}\right)_{T} &= - \frac{R T}{(v-b)^2}\left\{1-2a\left[R T(v+b)\left(\frac{v^2+2bv-b^2}{v^2-b^2}\right)^2\right]^{-1}\right\}\;.
    \end{align}
\end{subequations}

The speed of sound for general real fluids can be evaluated as
\begin{equation}
    \label{eqn:sos}
    c^2 = \left(\frac{\partial p}{\partial \rho}\right)_{s} = \frac{\gamma}{\rho \kappa_T}\,,
\end{equation}
where $\gamma$ is the specific heat ratio and $\kappa_T$ is the isothermal compressibility, which is defined as
\begin{equation}
    \kappa_T = -\frac{1}{v}\left(\frac{\partial v}{\partial p}\right)_{T}\,.
\end{equation}

The molecular transport properties, i.e., the dynamic viscosity and the thermal conductivity are evaluated using Chung's high-pressure method for dense fluids~\cite{chung1984, chung1988}.

\subsection{Numerical schemes}
A finite volume approach is utilized for the discretization of the system of equations, \cref{eqn:governingEqn}
\begin{equation}
    \label{eqn:fv}
    \frac{\partial \boldsymbol{U}}{\partial t} V_{cv} + \sum_f (\boldsymbol{F}^e_f - \boldsymbol{F}^v_f) A_f = \boldsymbol{S}\,,
\end{equation}
where 
\begin{equation}
    \boldsymbol{U} = \left[\rho, (\rho \boldsymbol{u})^T, \rho E \right]^T
\end{equation}
is the vector of conserved variables, $\boldsymbol{F}^e$ is the face-normal Euler flux vector, $\boldsymbol{F}^v$ is the face-normal viscous flux vector which corresponds to the right-hand side (RHS) of \cref{eqn:governingEqn}, $\boldsymbol{S}$ is the source term vector, $V_{cv}$ is the volume of the control volume, and $A_f$ is the face area.

A Strang-splitting scheme~\cite{strang1968construction} is applied in this study to separate the convection and diffusion operators. The convective flux is discretized using a sensor-based hybrid scheme in which a high-order, non-dissipative scheme is combined with a low-order, dissipative scheme to minimize the numerical dissipation~\cite{MaJCP2017, khalighi2011unstructured}. A central scheme which utilizes a fourth-order reconstruction on uniform mesh is used along with a second-order ENO scheme for the hybrid scheme. A density sensor~\cite{MaJCP2017, hickey2013large} is adopted in this study. Due to the large density gradients under transcritical conditions, an entropy-stable flux correction technique~\cite{MaJCP2017} is used to ensure the physical realizability of the numerical solutions and to dampen non-linear instabilities in the numerical schemes. A strong stability preserving 3rd-order Runge-Kutta (SSP-RK3) scheme \cite{gottlieb2001strong} is used for time integration.

For a fully conservative (FC) scheme, the Euler flux of faces of cell $i$ is evaluated as
\begin{equation}
    \boldsymbol{F}^e = \hat{\boldsymbol{F}} (\boldsymbol{U}_j)\;\; \text{for} \;\; j \in \boldsymbol{I}_i\,,
\end{equation}
where $\hat{\boldsymbol{F}}$ is the numerical flux, and $\boldsymbol{I}_i$ is the spatial stencil of cell $i$. The evaluation of numerical flux $\hat{\boldsymbol{F}}$ involves spatial reconstruction and flux calculations. The conservative variables are updated using \cref{eqn:fv}, after which the primitive variables are calculated using the updated conservative variables. Thermodynamic variables are updated using the specified EoS, and specifically the pressure is updated by \cref{eqn:eos}, which is generally an iterative process for real-fluid EoSs. However, due to the strong non-linearity inherent in the real-fluid EoS, it is well known that spurious pressure oscillations will be generated when a FC scheme is used~\cite{terashima2012approach, pantano2017oscillation, MaJCP2017, lacaze2017robust}. Therefore, a quasi-conservative scheme using a double-flux (DF) model~\cite{ma2014supercritical, MaJCP2017, ma2017numerical} is utilized in this study, and the performance of these two schemes will be assessed. In the DF scheme, the relation between the pressure and the internal energy is frozen in both space and time, which converts the local system to an equivalently calorically prefect gas system. This treatment removes not only the spurious pressure oscillations but also the oscillations in other physical quantities induced by the pressure oscillation. Only an outline of this method is presented here and the details are developed in Ma et al.~\cite{MaJCP2017}. In this model, an effective specific heat ratio or adiabatic exponent, and an effective reference energy value are used to relate the pressure and the internal energy, which are defined as
\begin{subequations}
    \label{eqn:gammaS}
    \begin{align}
        &\gamma^* = \frac{\rho c^2}{p}\,,\\
        &e_0^* = e - \frac{pv}{\gamma^* - 1}\,.
    \end{align}
\end{subequations}
The Euler flux for cell $i$ is evaluated as
\begin{equation}
    \boldsymbol{F}^e = \hat{\boldsymbol{F}} (\boldsymbol{U}^*_j)\;\; \text{for} \;\; j \in \boldsymbol{I}_i\,,
\end{equation}
where
\begin{equation}
     \boldsymbol{U}^* = \left[\rho, (\rho \boldsymbol{u})^T, \rho E^* \right]^T
\end{equation}
with
\begin{equation}
    (\rho E)^*_j = \frac{p_j}{\gamma^*_i - 1} + \rho_j e_{0,i}^* + \frac{1}{2} \rho_j \boldsymbol{u}_j \cdot \boldsymbol{u}_j\,.
\end{equation}
After the conservative variables are updated using \cref{eqn:fv}, the pressure is calculated as
\begin{equation}
    p_i = (\gamma^*_i - 1) \left[(\rho E)_i - \rho e_{0,i}^* - \frac{1}{2} \rho_i \boldsymbol{u}_i \cdot \boldsymbol{u}_i \right]\,,
\end{equation}
with other thermodynamic quantities updated using the pressure. Finally, after all the RK sub-steps, to ensure the thermodynamic consistency, the internal energy is updated from the primitive variables using \cref{eqn:eos}. Note that the formulation of the relation between energy and pressure described by \cref{eqn:gammaS} ensures that the characteristic speed of sound of the Euler system of \cref{eqn:governingEqn} to be equal to the thermodynamic speed of sound. The conservation error in total energy was shown to be small and converge to zero with increasing resolution~\cite{MaJCP2017}.

\subsection{Computational details}
\begin{figure}[!tb!]
    \centering
    \includegraphics[height=0.38\columnwidth,clip=]{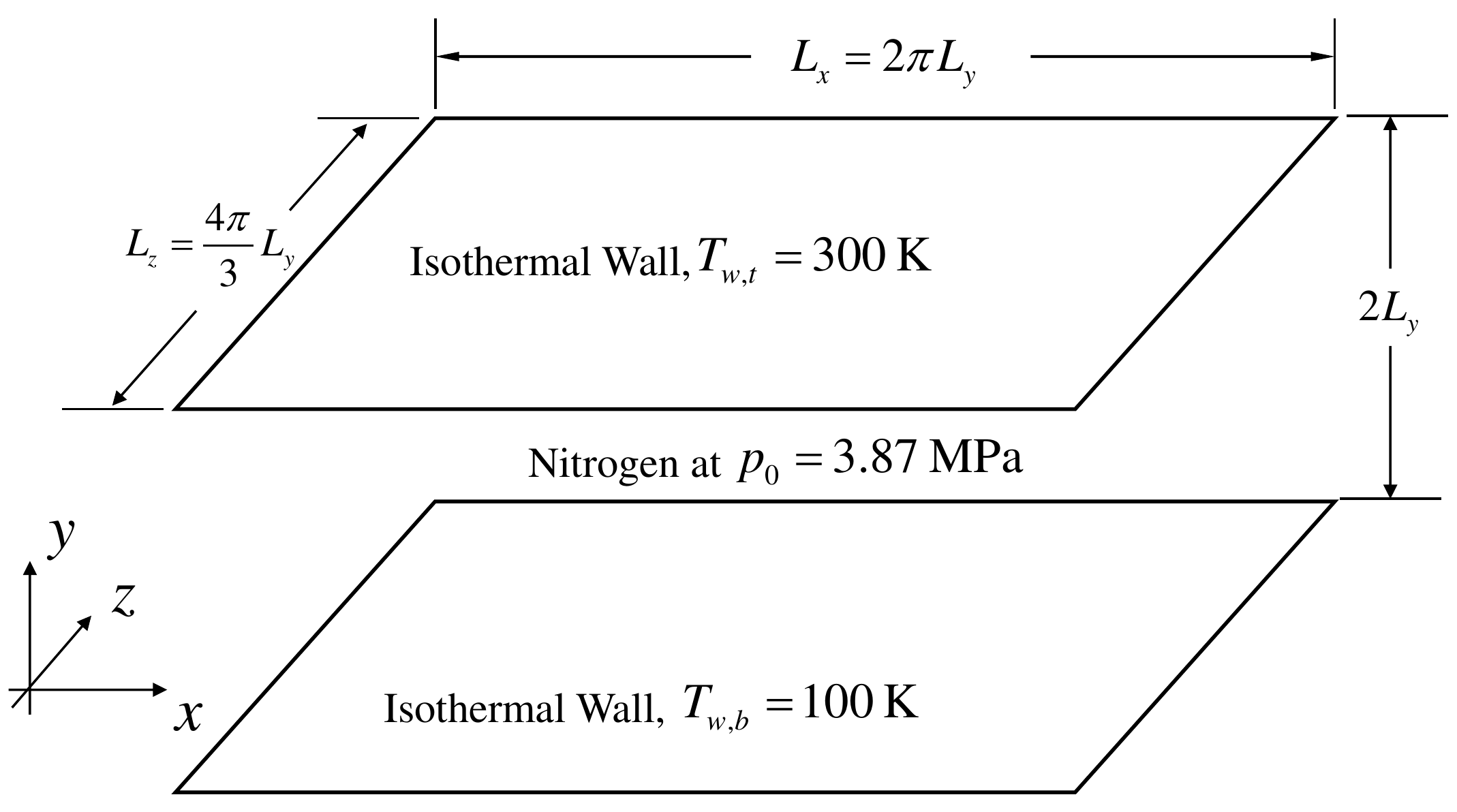}
    \caption{Schematic of the transcritical turbulent channel flow. The top and bottom walls are isothermal walls kept at 300~K and 100~K, respectively; $p_0$ is the bulk pressure.\label{fig:schematic}}
\end{figure}

The computational domain is schematically illustrated in \cref{fig:schematic}. The working fluid is nitrogen, whose critical pressure and critical temperature are $p_c=3.40$~MPa and $T_c=126.2$~K, respectively. The flow is at a bulk pressure of 3.87~MPa, corresponding to a reduced pressure ($p_r = p/p_c$) of 1.14, which is only slightly higher than the critical pressure. 
The flow is confined between two isothermal walls, which are kept at $T_{w,b} = 100$~K and $T_{w,t} = 300$~K, where $T$ is the temperature, and the subscripts $b$ and $t$ denote `bottom' and `top', respectively. The reduced temperature ($T_r = T/T_c$) is 0.79 at the bottom cooled wall and 2.38 at the top heated wall.
The periodic channel is of size $L_x \times 2L_y \times L_z$ with $L_x/L_y=2\pi$, $L_z/L_y=4\pi/3$ and half-channel height of $L_y = 0.09$~mm, where $x$, $y$ and $z$ are the streamwise, wall-normal and spanwise directions, respectively. The wall-normal coordinate extends from $y=-L_y$ to $y=L_y$. A constant flow rate is enforced and the bulk velocity, defined as $\widetilde{u}_0=\int \rho u dV/\int \rho dV$, is 27.3~m/s, where the integration is over the entire channel, $\rho$ is the fluid density, and $u$ is the streamwise velocity. 

The finite-volume compressible code CharLES$^{\rm{x}}$ is used in this study.  
This code has been extensively used for turbulent flow calculations \cite{hickey2013large, larsson2015incipient, wu2017mvp, lv2017underresolved}. Here we only briefly summarize the main features of the code, and further details can be found in \cite{khalighi2011unstructured}, \cite{MaJCP2017}, and  references therein. 
The flux reconstruction uses a central scheme where fourth-order accuracy is obtained on uniform mesh. A sensor-based hybrid central-ENO scheme is used to capture flows with large density gradients and to minimize the numerical dissipation while stabilizing the simulation. For regions where the density ratio between the reconstructed face value and the neighboring cells exceeds 25\%, a second-order ENO reconstruction is used on the left- and right-biased face values, followed by an HLLC Riemann flux computation. An entropy-stable double-flux model, developed for transcritical flows~\cite{MaJCP2017}, is employed to prevent spurious pressure oscillations and to ensure the physical realizability of numerical solutions. 
A Strang-splitting scheme~\cite{strang1968construction} is employed to separate the convection operator from the remaining operators of the system. A strong stability preserving third-order Runge-Kutta scheme~\cite{gottlieb2001strong} is used for time integration.
Equation \eqref{eqn:preos} is used as the constitutive equation, and the molecular transport properties, including the dynamic viscosity and the thermal conductivity, are evaluated according to Chung's model for high-pressure fluids~\cite{chung1984applications,chung1988generalized}.

For the present study, a structured grid is used, and the mesh is of size $N_x\times N_y\times N_z=384\times256\times384$, with uniform grid spacings in the streamwise and spanwise directions. The grid resolutions are $\Delta x^+=7.0$, $\Delta y^+_\text{min}=0.29$, $\Delta y^+_\text{max}=6.7$, $\Delta z^+=4.7$ based on the wall units at the bottom cooled wall, and $\Delta x^+=4.8$, $\Delta y^+_\text{min}=0.20$, $\Delta y^+_\text{max}=4.6$, $\Delta z^+=3.2$ based on the wall units at the top hot wall.
The friction Reynolds number $Re_\tau=L_y \rho_w u_\tau/\mu_w$ is $Re_{\tau,b}=430$ and $Re_{\tau,t}=300$ based on wall units at the bottom and top walls, respectively. 
DNS of incompressible channel flows typically require grid resolutions $\Delta x^+\approx 10$, $\Delta y^+_{min}\approx 0.5$, $\Delta y^+_{max}\approx 10$, $\Delta z\approx 10$ \cite{moser1999direct, lozano2014effect, lee2015direct}.
We have used a slightly finer grid for the transcritical flow calculation here following Kawai~\cite{kawai2016direct}. Because of the use of a fine grid, the flow is well resolved and the ENO scheme is active on less than $0.06\%$ of the cell faces. Simulations are advanced in time at a unity CFL number. After the flow reaches a statistically stationary state, we average across the homogeneous directions and over six flow-through times to obtain fully converged statistics, where one flow-through is defined as $t_f = L_x/\widetilde{u}_0$.

\section{Results and Discussion}\label{sec:results}

\subsection{Favre-averaged and Reynolds-averaged quantities}\label{sect:aver}
\begin{figure}[!t!]
    \centering
    \includegraphics[width=0.38\columnwidth]{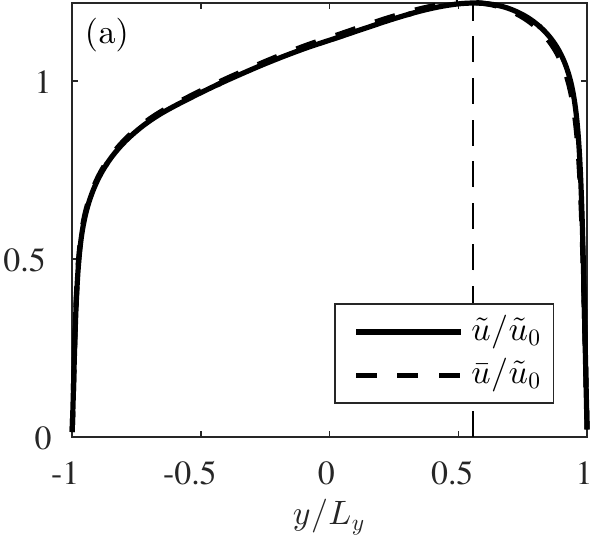}\quad\quad
    \includegraphics[width=0.38\columnwidth]{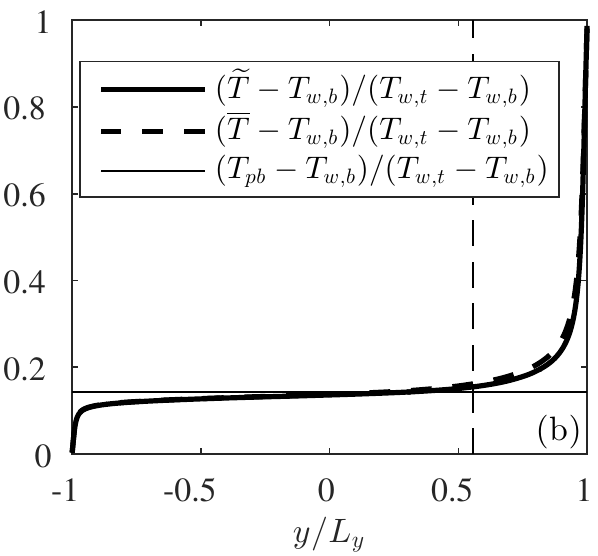}
    \caption{Favre- and Reynolds-averaged (a) velocity and (b) temperature as a function of the wall-normal coordinate. $T_{pb}$ is the pseudo-boiling temperature, where the specific heat capacity peaks. The dashed line is at wall-normal coordinate $y=y_\delta$, where $\tilde{u}$ is at its maximum.}
    \label{fig:umTm}
\end{figure}

As we consider variable-property fluids, we start our discussion by examining Favre- and Reynolds-averaged quantities. \Cref{fig:umTm} shows the Favre- and Reynolds-averaged velocity and temperature as a function of the wall-normal coordinate. Here the Reynolds average of a quantity $\phi$ is the conventional ensemble average, which is denoted as $\overline{\phi}$, and the Favre average is defined as $\widetilde{\phi} = \overline{\rho \phi} / \overline{\rho}$. Since the Favre and the Reynolds averages differ by $\overline{\phi^\prime \rho^\prime}$, which equals Cor$_{\rho^\prime \phi^\prime}$std$(\rho^\prime)$std$(\phi^\prime)$~\cite{SMITS_DUSSAUGE_BOOK2006}, this similarity can only be a consequence of Cor$_{\rho^\prime \phi^\prime}\ll1$ or std$({\rho^\prime})$std$({\phi^\prime})\ll\overline{\rho}\overline{\phi}$, where the superscript $\prime$ indicates a fluctuating quantity and Cor$_{\rho^\prime \phi^\prime}$ is the correlation between the  fluctuations of $\rho$ and $\phi$. Here, $\widetilde{\phi}\approx\overline{\phi}$ is because std$({\rho^\prime})$std$({\phi^\prime})\ll\overline{\rho}\overline{\phi}$, where $\phi=\{u,T\}$. The streamwise velocity fluctuations and the temperature fluctuations are in fact both well correlated with the density fluctuations (the correlation Cor$_{\phi^\prime \rho^\prime}$ is $\approx 0.9$ in the near wall region and is $\approx 0.5$ in the bulk region). 

\Cref{fig:umTm}(a) shows an asymmetric velocity profile, with the momentum boundary layer near the top heated wall being thinner than that near the bottom cooled wall. The wall-normal height at which $\overline{u}$ attains its maximum is defined as $y_{\delta}$, which takes the value of $0.56L_y$ (see dashed line).
The top and bottom wall boundary layer heights are thus determined by taking the difference between $y_\delta$ and the $y$-coordinate of the two walls. Instead of using $y_\delta$, one can also use the height at which $\overline{u^\prime v^\prime}=0$, which yields essentially the same wall-normal coordinate. The same asymmetry is found in the temperature profile. Moreover, we notice that the temperature in the bulk of the channel is fairly close to the pseudo-boiling temperature, $T_{pb}=128.7$~K. 

\begin{figure}[!t!]
    \centering
    \includegraphics[width=0.43\columnwidth]{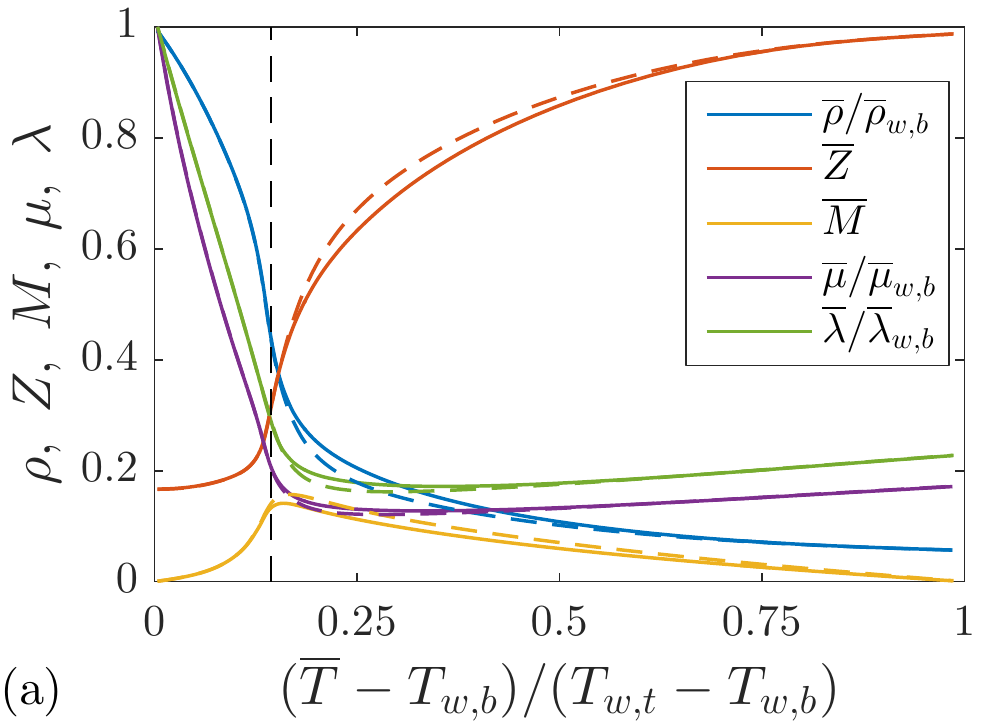}\quad\quad
    \includegraphics[width=0.428\columnwidth]{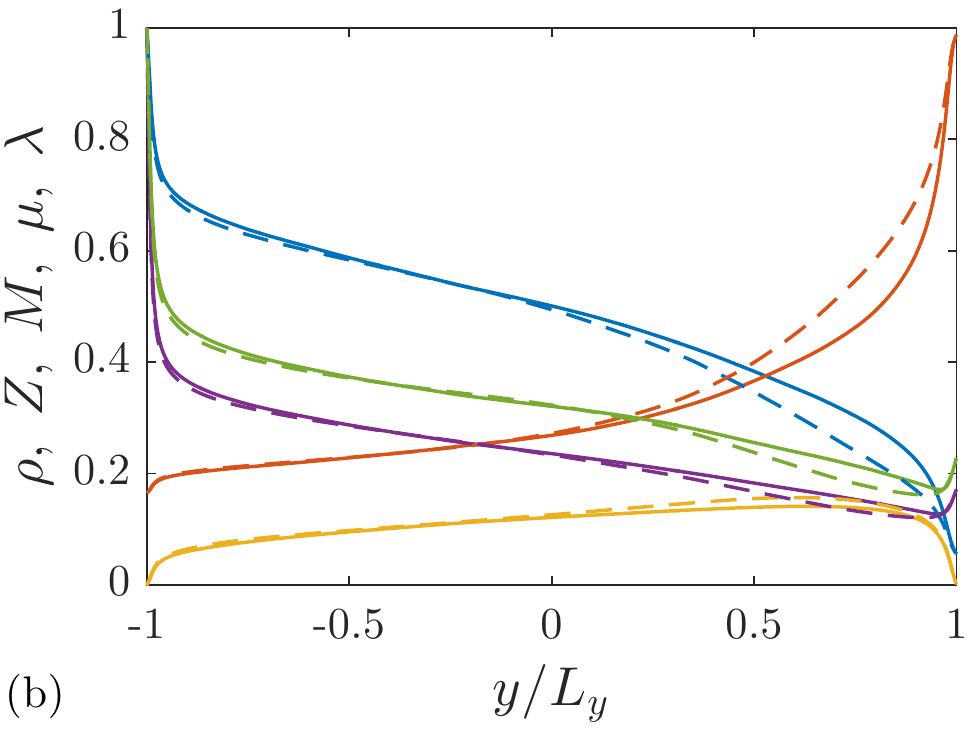}
    \caption{Ensemble-averaged density ($\rho$), compressibility factor ($Z$), Mach number ($M$), Prandtl number (Pr) and specific heat capacity ($c_p$) as (a) a function of the normalized mean temperature and (b) a function of the wall-normal coordinate. Prandtl number and specific heat capacity are normalized by their maximum values, which are 6.93 and $1.12\times10^4$~J/(kg$\cdot$K), respectively. Density is normalized by the value at the bottom wall, which is 785~kg/m$^3$. Results from both FC (solid) and DF (dashed) schemes are shown.}
    \label{fig:thermal}
\end{figure}

\begin{figure}
    \centering
    \includegraphics[width=0.54\columnwidth]{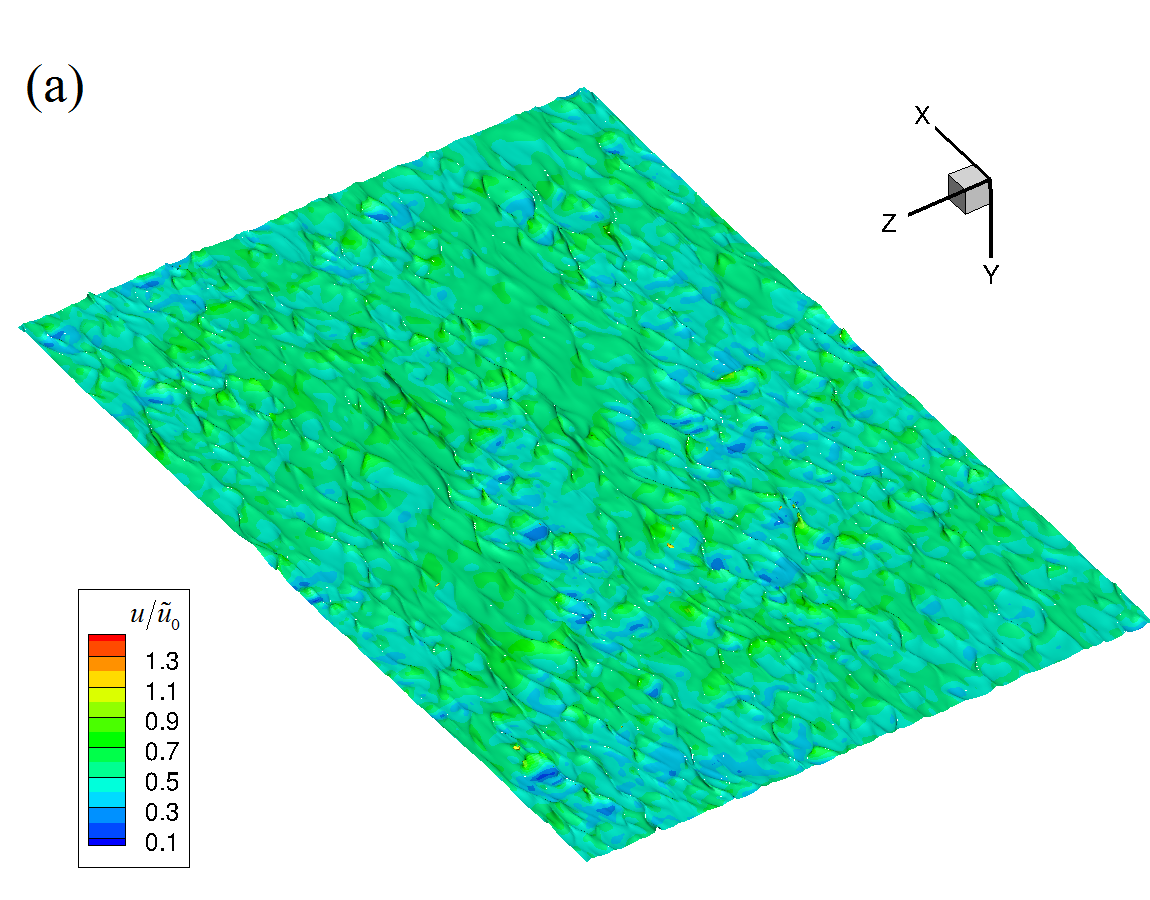}\\[1em]
    \includegraphics[width=0.54\columnwidth]{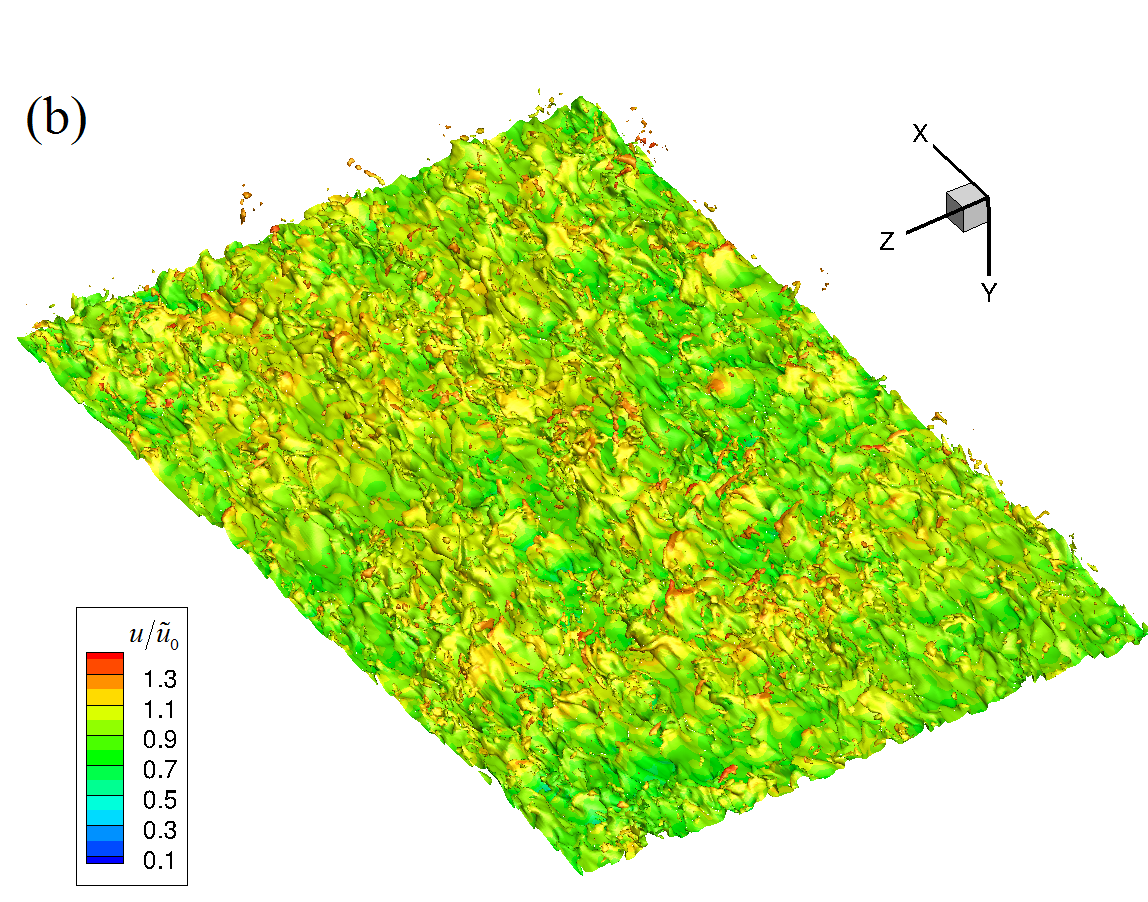}\\[1em]
    \includegraphics[width=0.54\columnwidth]{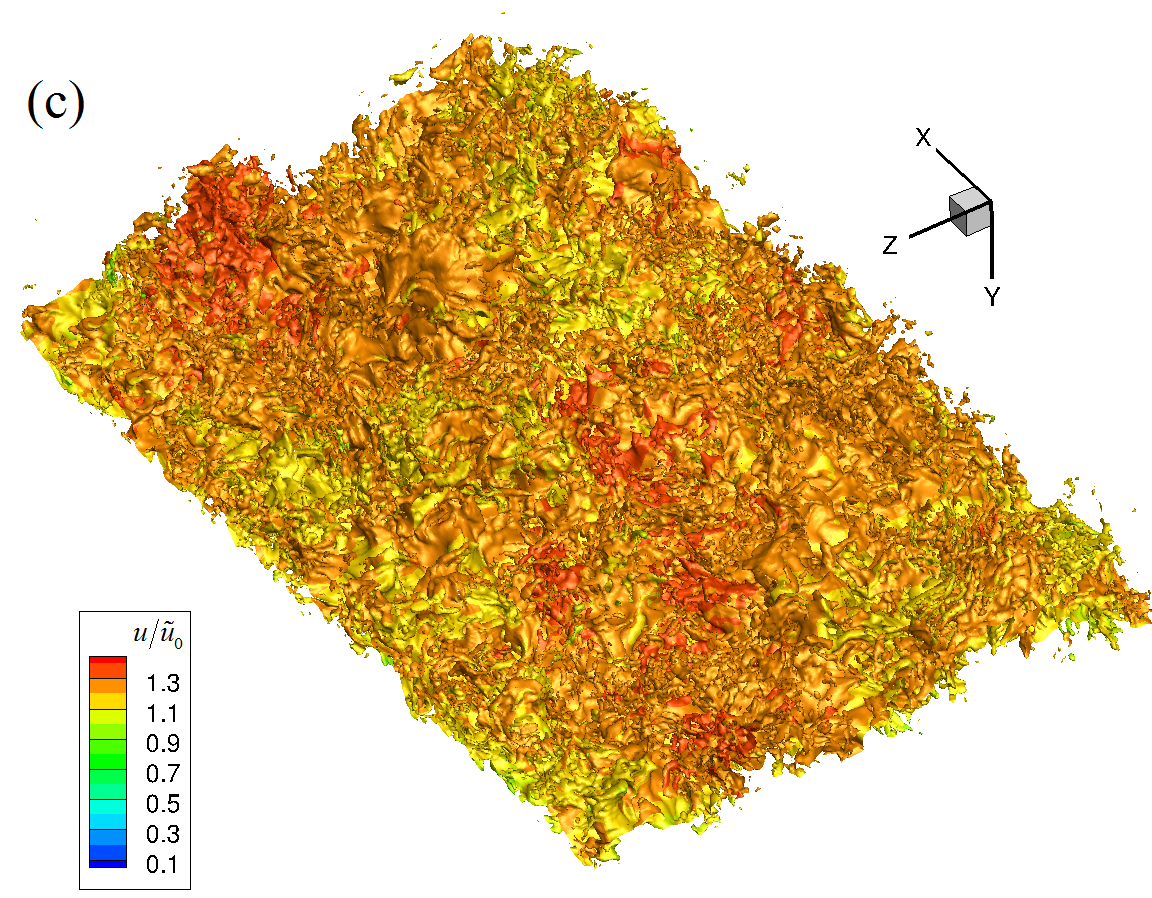}
    \caption{Isosurfaces of constant density, colored by streamwise velocity. (a) $\rho=60$~kg/m$^3$ ($y_s/L_y=0.98$), (b) $\rho=100$~kg/m$^3$ ($y_s/L_y=0.9$) and (c) $\rho=200$~kg/m$^3$ ($y_s/L_y=0.6$). The fluid density at top and bottom walls is 44 kg/m$^3$ and 785 kg/m$^3$, respectively;
    $y_s$ denotes the mean location of the isosurface of density. Results shown are from the DF scheme.}
    \label{fig:isocontour}
\end{figure}

Next, we examine the thermal properties. \Cref{fig:thermal} shows ensemble-averaged values of density ($\rho$), compressibility factor ($Z$), Mach number ($M$), dynamic viscosity ($\mu$) and heat conductivity ($\lambda$) as a function of mean temperature $\overline{T}$ and wall-normal coordinate.
Near the top heated wall, the fluid can already be approximated as ideal gas ($\overline{Z}$ approaches unity).
The density and compressibility change drastically over a few degrees of Kelvin near the pseudo-boiling temperature $T_{pb}$.
Because $\overline{T(y)}$ is close to the pseudo-boiling temperature $T_{pb}$ in the bulk region, the wall-normal gradients of these quantities in the bulk region are comparably moderate.
Density, dynamic viscosity and heat conductivity change drastically near the two walls as a function of the wall-normal distance.
Last, the Mach number is everywhere below $0.16$, so that this configuration corresponds to the low-speed flow regime.

In \cref{fig:thermal}, results using both the FC and DF schemes are shown in solid and dashed lines, respectively. It can be seen that results with the DF scheme reasonably agrees with the FC scheme. However, noticeable difference can be observed between the results from the two schemes, especially near the top heated wall, where the temperature is above the pseudo-boiling value. Similar difference was observed in previous studies on turbulent flat-plate boundary layer calculations when both FC and QC schemes were employed~\cite{kawai2015robust}. This difference can be attributed to the energy conservation error introduced by the DF scheme, even with the relatively fine grid resolution that the present study adopt.

\subsection{Instantaneous flow field}\label{sect:inst}
We continue our analysis by examining the instantaneous flow field and discuss statistical results pertaining to the near-wall structure.
\Cref{fig:isocontour} shows instantaneous isosurfaces of $\rho = \{60, 100, 200\}$ kg/m$^3$ from results using the DF scheme. 
The fluid density is 44 kg/m$^3$ and 785 kg/m$^3$ at the top and bottom walls, respectively. 
\Cref{fig:isocontour}(a)-(c) are at increasing distances from the top heated wall. 
Footprints of $\Lambda$-vortices can be discerned in \cref{fig:isocontour}(a). 
Further into the bulk region, at $\rho=100$~kg/m$^3$, the near-wall vortices break down, leading to turbulent spots among comparably quiescent regions~\cite{wu2017transitional}.
At $\rho=200$~kg/m$^3$, the isosurface becomes highly corrugated, indicating strong mixing. 
We also note that the density and the velocity in this particular flow are well-correlated, and there is barely any variation in the coloring of each isosurface.

\begin{figure}[!b!]
    \centering
    \includegraphics[width=0.49\columnwidth]{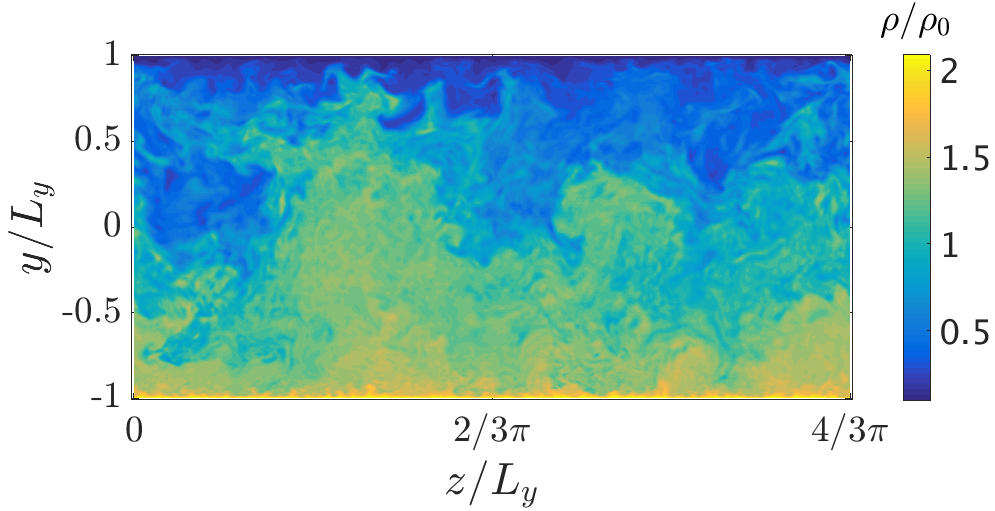}
    \includegraphics[width=0.49\columnwidth]{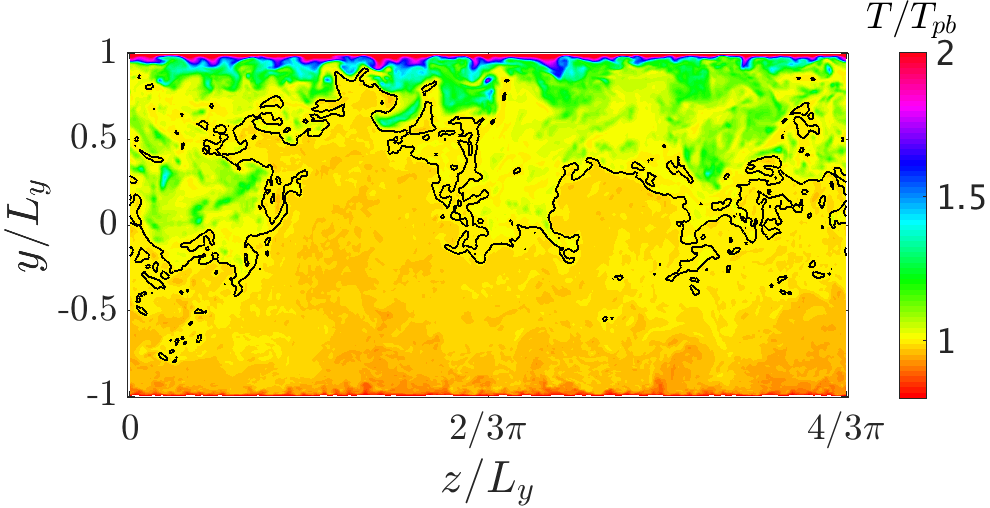}
    \caption{Instantaneous contours on an $z$\textendash$y$ plane for density ($\rho_0$ is the volume-averaged bulk density) (left) and temperature (the black line indicates $T/T_{pb}=1$) (right) using the DF scheme.}\label{fig:instx}
\end{figure}

\begin{figure}[!b!]
    \centering
    \includegraphics[width=0.49\columnwidth]{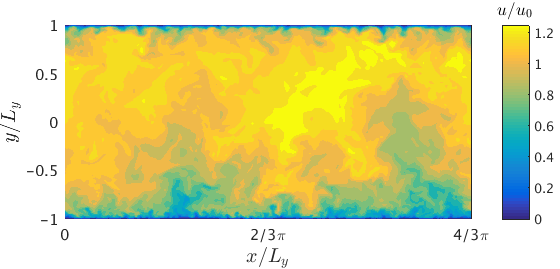}
    \includegraphics[width=0.49\columnwidth]{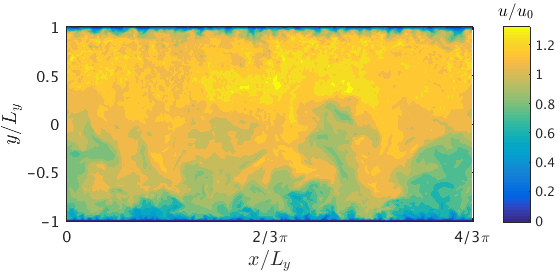}
    \caption{Instantaneous contours on an $z$\textendash$y$ plane for streamwise velocity using both DF (left) and FC (right) schemes.}
    \label{fig:ux}
\end{figure}

\Cref{fig:instx,fig:ux} shows instantaneous contours of the fluid density, temperature, and streamwise velocity on a $z$\textendash$y$ plane. Violent density fluctuations can be seen in \cref{fig:instx}, with intrusions of high-density fluid into fluid of lower density and \emph{vice versa}. Also, it is apparent from \cref{fig:ux} that the momentum boundary layer is significantly thinner near the top heated wall. Results for streamwise velocity are compared between FC and DF schemes in \cref{fig:ux}. It can clearly be seen that the spurious oscillations in the velocity field are present for the FC scheme, which are induced by the spurious pressure oscillations~\cite{terashima2012approach,MaJCP2017}. The region with strong spurious oscillations is near the top heated wall, which is in consistent with the results from \cref{fig:thermal}, which serves as a good example of the dilemma between energy conservation and pressure equilibrium~\cite{karni1994multicomponent, abgrall1996prevent}.

\subsection{Attached eddy model}\label{sect:AEH}
\begin{figure}[!b!]
    \centering
    \includegraphics[width=0.4\columnwidth]{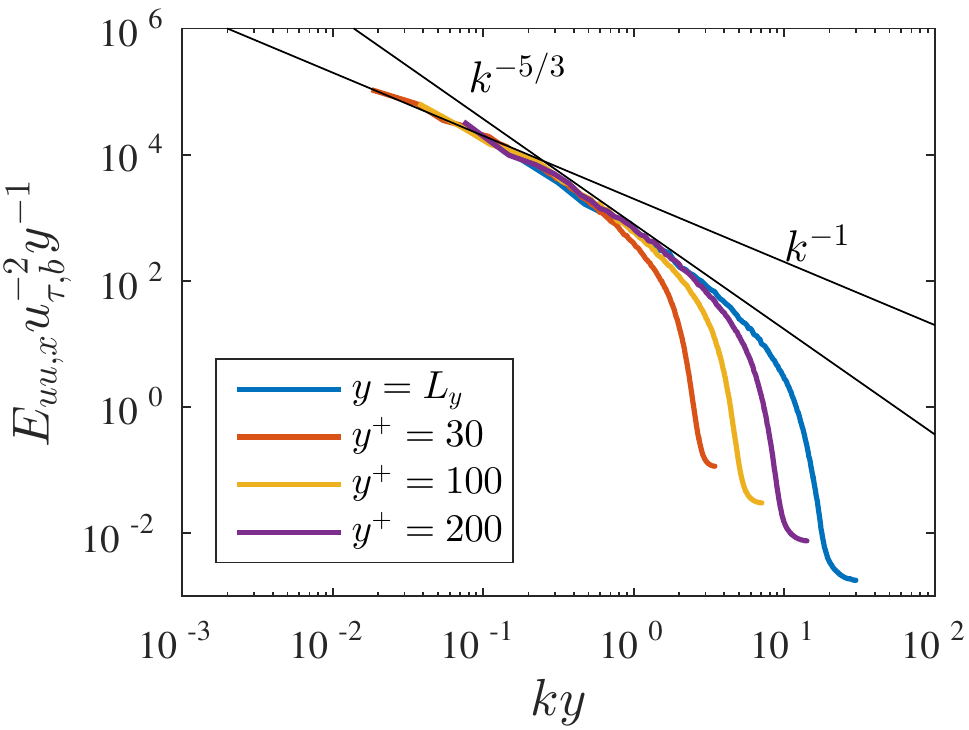}\quad\quad
    \includegraphics[width=0.4\columnwidth]{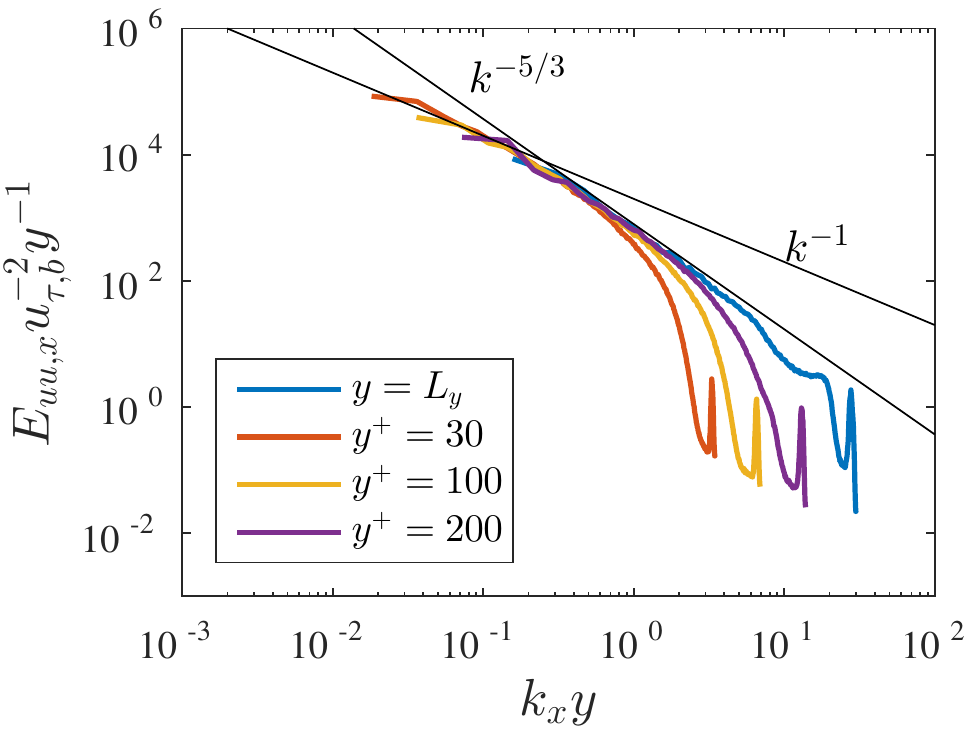}
    \caption{Energy spectra of the streamwise velocity fluctuation $u^\prime$ as a function of the streamwise wave number at various wall-normal distances from the bottom cooled wall using both DF (left) and FC (right) schemes. }
    \label{fig:specxb}
\end{figure}

In this subsection, we consider the attached eddy hypothesis of Townsend~\cite{townsend1980structure}. 
This hypothesis considers high-Reynolds-number wall-bounded flows as collections of wall-attached, self-similar, space-filling eddies, whose sizes scale with their distance from the wall.
The attached eddy model has been tested extensively, and has been proven to be quite useful \cite{meneveau2013generalized, yang2016extended, yang2016moment, woodcock2015statistical}.
While there are other statistics that can be deduced from the attached eddy model, the $k^{-1}$ spectrum of the streamwise velocity fluctuations as related to the constant momentum flux region is at the center of the attached eddy formalism \cite{nickels2005evidence}. \Cref{fig:specxb} shows the energy spectrum of $u^\prime$ near the bottom wall. In this region, both the $k^{-1}$ law and $k^{-5/3}$ law survive in the spectra and the spectra collapse using regular normalizations. 
A similar behavior cannot be found near the top wall, and variations in fluid density must be accounted for. 

\begin{figure}[!b!]
    \centering
    \includegraphics[width=0.4\columnwidth]{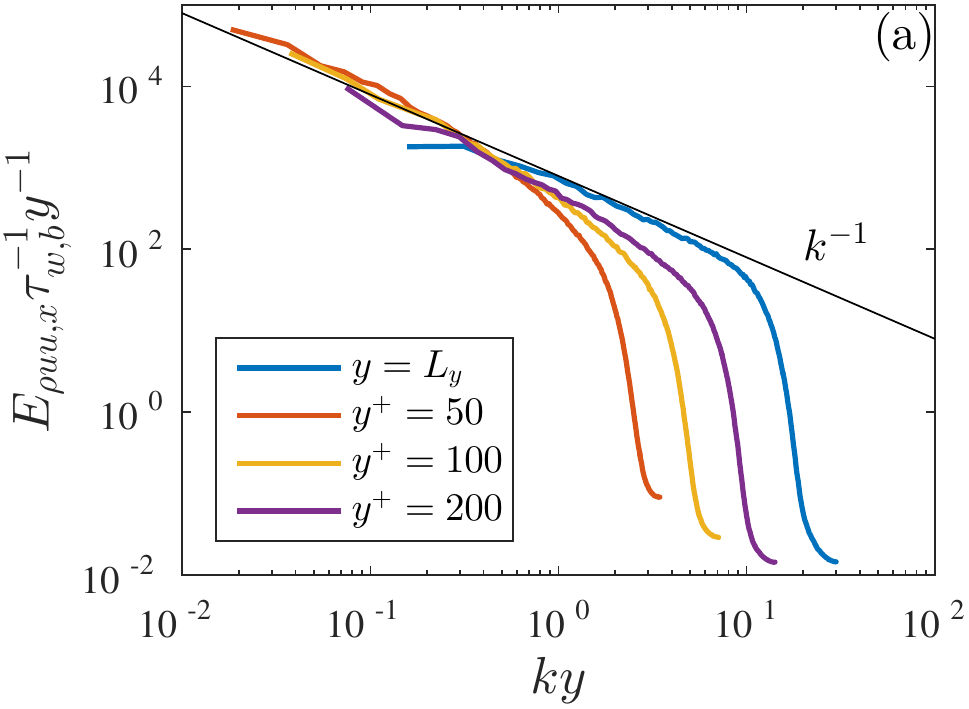}\quad\quad
    \includegraphics[width=0.405\columnwidth]{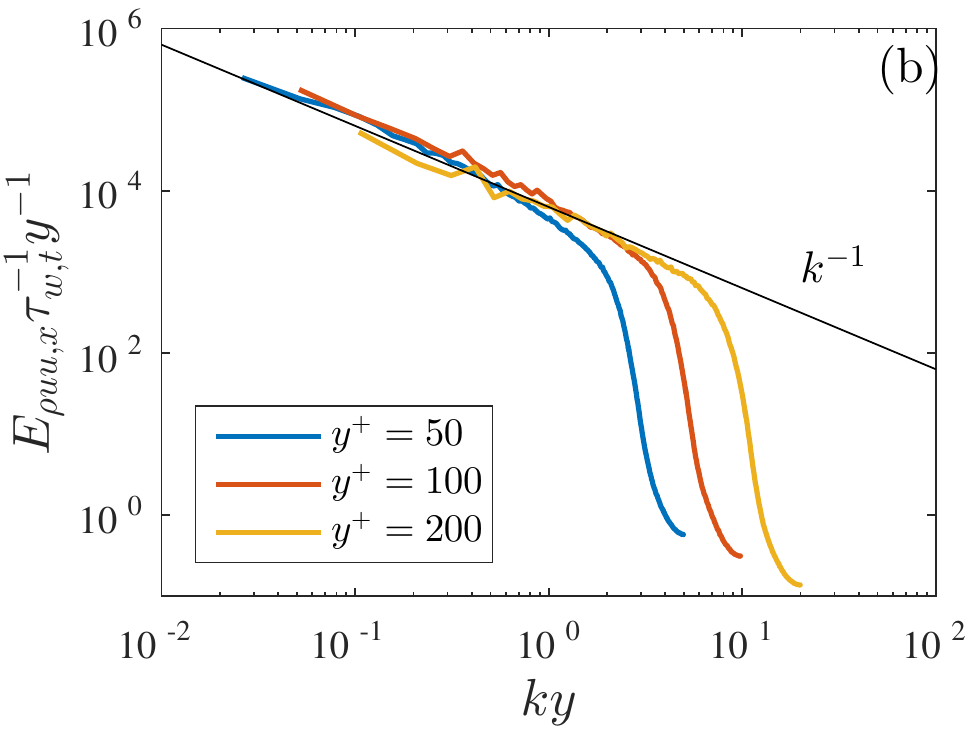}
    \caption{Energy spectra of $(\sqrt{\rho}u)^\prime$ as a function of the streamwise wavenumber at different wall-normal distances from (a) the bottom cooled wall and (b) the top heated wall using the DF sheme.}
    \label{fig:specrx}
\end{figure}

\begin{figure}[!t!]
    \centering
    \includegraphics[width=0.4\columnwidth]{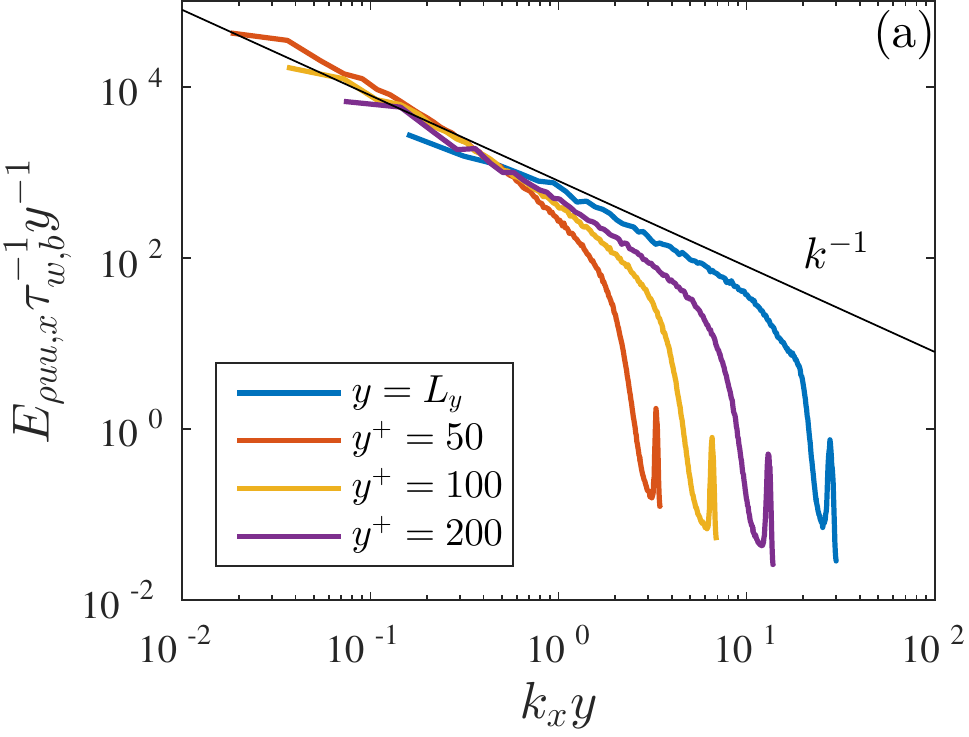}\quad\quad
    \includegraphics[width=0.405\columnwidth]{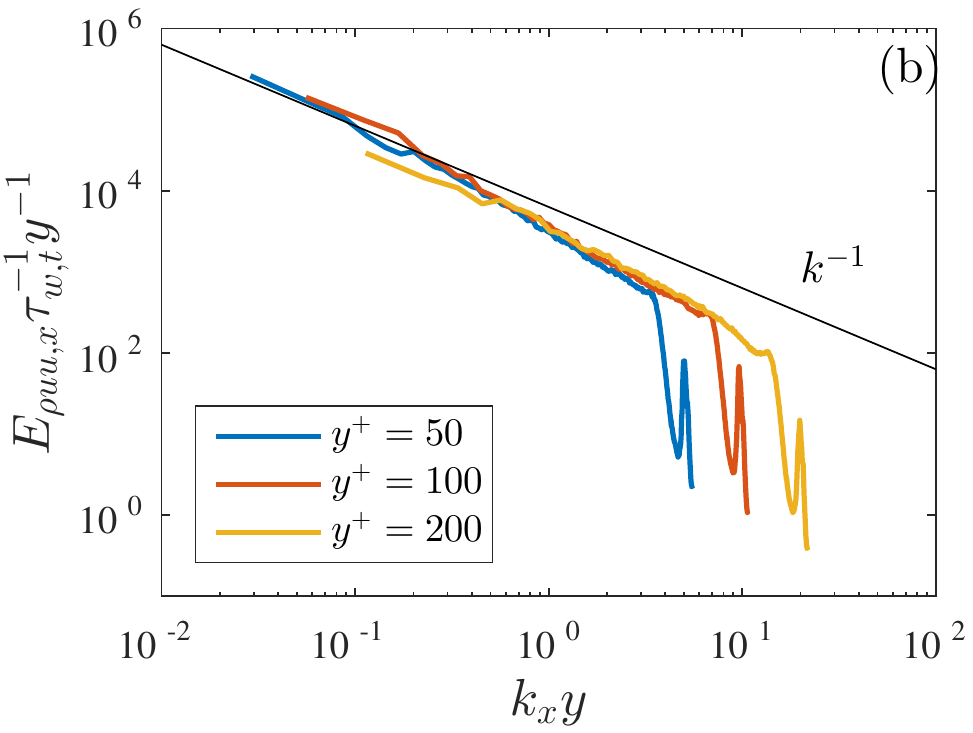}
    \caption{Energy spectra of $(\sqrt{\rho}u)^\prime$ as a function of the streamwise wavenumber at different wall-normal distances from (a) the bottom cooled wall and (b) the top heated wall using the FC scheme.}
    \label{fig:specrxFC}
\end{figure}

\Cref{fig:specrx,fig:specrxFC} shows energy spectra of $(\sqrt{\rho}u)^\prime$ as a function of the streamwise wave number at several wall-normal distances from the bottom cooled wall and the top heated wall. The spectra are normalized with $\tau_w$, which equals the momentum flux in the inertial range, and the wall-normal distance, which is nominally the only relevant length scale in the log region. 
According to \cref{fig:specrx}, the energy spectra follow the $k^{-1}$ scaling across a wide range of scales and decays rapidly at small scales. 
Interestingly, the $k^{-5/3}$ spectra is absent. 
Although local isotropy is formally only expected at infinite Reynolds numbers, it is still rather surprising that the $-5/3$ scaling could not be found here. 

Results in \cref{fig:specxb,fig:specrx,fig:specrxFC} are presented for both FC and DF schemes, which qualitatively agree well with each other. However, it can clearly be seen that due to the spurious pressure oscillations, which contaminate the velocity field, spikes are present in the high-wavenumber region for the FC scheme. In contrast, the spectrum calculated from results with the DF scheme show smooth tails at high wavenumbers.

\begin{figure}[!t!]
    \centering
    \includegraphics[width=0.48\columnwidth]{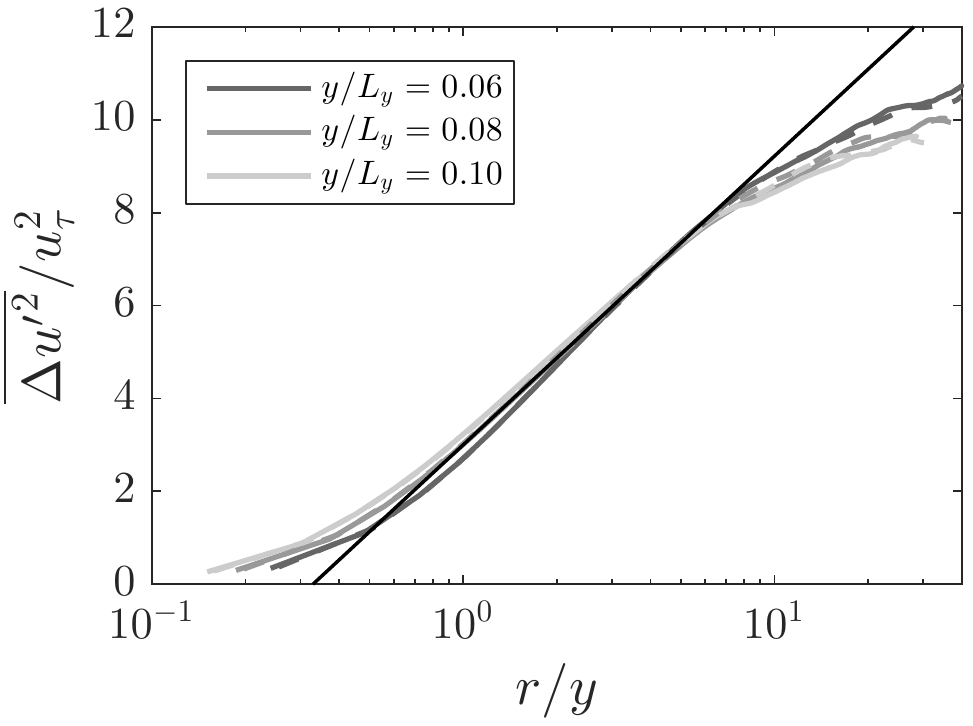}
    \caption{Streamwise structure function $\overline{\Delta {u^\prime}^2}/u_\tau^2$ as functions of the two-point displacement distance. 
    The FC results and the DF results are shown using solid and dashed lines, respectively.}
    \label{fig:str}
\end{figure}

The attached-eddy model admits scaling laws including 
\begin{equation}
    \overline{u^{\prime2}}\sim \log(\delta/y)\,, ~~~\overline{w^{\prime2}}\sim \log(\delta/y)\,,
    ~~~\overline{\Delta {u^\prime}^2}=\overline{(u^\prime(x)-u^\prime(x+r))^2}\sim \log(r/y)\,,\label{eq:logAEH}
\end{equation} 
where $\delta$ is an outer length scale and $r$ is a two-point displacement.
All scalings have gained considerable empirical support from both laboratory experiments and numerical simulations \cite{de2015scaling, stevens2014large, hultmark2012turbulent, lee2015direct}.
While the logarithmic scaling of the variance of the streamwise velocity fluctuations is only observed at high Reynolds numbers, the scaling of the spanwise velocity fluctuations can be found at moderate Reynolds numbers (in a region $100\lesssim y^+$, $y/\delta\lesssim 0.5$). 
Figure \ref{fig:str} shows the streamwise structure functions as functions $\overline{\Delta {u^\prime}^2}$ as functions of the two-point distance at sevearl wall-normal locations.
A logarithmic behavior is found and the data collapse as a function of $r/y$.

\section{Conclusions}\label{sec:conclusions}
In this work, DNS of a channel flow at transcritical conditions are conducted using both fully (FC) and quasi-conservative (QC) schemes. 
A temperature difference of 200~K is considered between the two isothermal walls for nitrogen at a bulk pressure only slightly higher than the critical pressure. The density changes by a factor of 18 in the flow field. 
A grid resolution slightly finer than that typically required for DNS of incompressible channel flows is adopted in this study.
The fluid temperature changes drastically near the two walls, however, in the bulk region, $\overline{T}\approx T_{pb}$.
As a result of heating, the boundary layer is thinner near the top heated wall (and at a lower Reynolds number). 
Comparing the results from both FC and QC schemes, noticeable difference in the mean values of different thermodynamic quantities is observed near the top heated wall, which is attributed to the energy conservation error present in the QC scheme. Spurious oscillations in streamwise velocity calculated by the FC scheme are also observed in this region, indicating a dilemma between energy conservation and pressure equilibrium.
Using the DNS data, we have examined the usefulness of the attached eddy model in the context of turbulence of variable thermodynamic properties. 
According to the DNS, the $k^{-1}$ spectrum survives, but a range of scales within which the flow is locally isotropic could not be found, at least not near the top heated wall. 
In addition to the $k^{-1}$ spectrum, we have also examined the streamwise structure function. 
An appreciable extent of the logarithmic range is found, suggesting that the physical picture proposed by previous studies~\cite{townsend1980structure, woodcock2015statistical, yang2016hierarchical} is still very much valid for flows at transcritical conditions.

\section*{Acknowledgments}
Financial support through ARL with award \#W911NF-16-2-0170, NASA with award \#NNX15AV04A, and AFOSR with award \#1194592-1-TAAHO are gratefully acknowledged. Peter C. Ma would also like to thank Dr. Daniel T. Banuti and Hao Wu for their help on this paper. Xiang I. A. Yang would like to thank P. Moin for his support.
Resources supporting this work were provided by the NASA High-End Computing (HEC) Program through the NASA Advanced Supercomputing (NAS) Division at Ames Research Center.

\bibliographystyle{aiaa}
\bibliography{references}

\end{document}

%% file: abbrevs.tex
\makeatletter
\newcommand\rlarrows{\mathop{\operator@font \rightleftarrows}\nolimits}
\makeatother
\DeclareFontFamily{OT1}{pzc}{}
\DeclareFontShape{OT1}{pzc}{m}{it}{<-> s * [1.10] pzcmi7t}{}
\DeclareMathAlphabet{\mathpzc}{OT1}{pzc}{m}{it}

 \def\0vec{{\mbox{\boldmath$0$}}}